\documentclass[acus]{JAC2000}

\usepackage{graphicx}

\begin{document}

\title{Measuring the Phases of $G_E$ and $G_M$ of the Nucleon }

\author{Stephen Rock, University of Massachusetts, Amherst, Mass. 01003, USA}

\maketitle
\begin{abstract}
 The nucleon electromagnetic form factors 
 $G_E$ and $G_M$ are complex quantities in the time-like region. The absolute
values can be determined by measuring the angular distribution of the nucleons
in $e^+ e^- \rightarrow N \bar{N}$. The complex phase can only be determined
by measuring one or more polarizations of the initial or final state. For PEP-N,
we can use unpolarized $e^+$ and $e^-$ and measure the polarization of one of 
the outgoing nucleons. 
\end{abstract}

\section{INTRODUCTION} 
 The Electromagnetic Form Factors of the nucleons, $G_E(Q^2)$ and $G_M(Q^2)$, 
depend on $Q^2$. 

In the space-like region they are relatively real. They have been measured 
using $e^- +N \rightarrow e^- +N$ elastic scattering where $Q^2= |q^2|$ is the 
is the absolute value of the four-momentum transfer from the incoming lepton to the 
nucleon. The techniques used are: 
\begin{enumerate}
\item Rosenbluth separation: Measurements made at at least two angles and fixed $Q^2$.\\
$d\sigma/d\Omega \propto{ (G_E(Q^2)^2 + \tau G_M(Q^2)^2)/( 1+\tau)} + 2\tau G_M(Q^2)^2 \sin^2(\theta/2)$
\item Polarized Beam and polarized Target 
 \begin{enumerate}
 \item target polarization in scattering plane $\perp$ $q$ vector: \\
 $ A_\perp = -P_eP_t {\sqrt{2\tau (1-\epsilon)}G_EG_M /
 (\epsilon G_E^2 +\tau G_M^2)}$
 \item Target polarization  $\parallel$ to $q$ vector:\\
 $ A_\parallel = -P_eP_t{ \sqrt{1-\epsilon^2}\tau G_M^2 /
 (\epsilon G_E^2 +\tau G_M^2)}$
 \end{enumerate}
\item Polarized beam and polarization of recoil nucleon: 
 \begin{enumerate}
 \item Recoil polarization in scattering plane $\perp$ $q$ vector: \\
  $ P_x = -P_e {\sqrt{2\tau (1-\epsilon)}G_EG_M/( \epsilon G_E^2 +\tau G_M^2)}$
 \item Recoil polarization  $\parallel$ to $q$ vector:\\
  $ P_z = P_e{ \sqrt{1-\epsilon^2}\tau G_M^2 /(\epsilon G_E^2 +\tau G_M^2)}$
 \end{enumerate}
\end{enumerate}
The first method gives only the absolute value of the Form Factors.
Combining   measurements 2a and 2b gives the relative sign thru:
${G_E/G_M}= \sqrt{\tau(1+\epsilon)/2\epsilon}\cdot{A_\perp/ A_\parallel}$.
Combining measurements 3a and 3b gives the relative sign thru:
${G_E/ G_M}= - \sqrt{\tau(1+\epsilon)/2\epsilon}\cdot{P_x/P_z}$.
All three methods have been used in the space like region for both the proton
and the neutron. \cite{Petratos}

In the time-like region $q^2$ is positive and the Form Factors are complex, so it is
necessary to measure  $|G_E|$, $|G_M|$ and the  Phase Difference. 
These can be determined using
the process $e^+ e^- \rightarrow N \bar{N}$. The momentum transfer $q^2 =s$ where
$s$ is the square of the center of mass energy. 
The  $N \bar{N}$ are in an L=0 ($G_s$) of L=1 ($G_d$) state with $G_M = G_s - G_d$
and $ G_E = \sqrt{s}/(2M)\cdot G_s +2G_d$. At threshold $G_d=0$ and thus
 $G_M(4M^2) = G_E(4M^2)$ and $Im[G_EG_M^*]=0$.
The VMD model of Dubnicka, Dubnickova and Strizenec \cite{DDS} gives predictions of 
the magnitude and phases of the Form Factor and predict significant non zero
phase difference in the PEP-N energy region.

The following experiments are possible.
\begin{enumerate}
 \item Unpolarized beam particles.
 \begin{enumerate}
 \item Rosenbluth separation: Measurements made at at least two angles and fixed $Q^2$.\\
$  d\sigma/d\Omega ={\alpha^2\sqrt{1-4M^2/x}/(4s)} 
  [|G_E(s)|^2\sin^2(\theta)/\tau +|G_M(s)|^2(1+\cos^2\theta)]$
 \item Recoil polarization  $\perp$ to scattering plane: \\
 $  P_y = -{\sin(2\theta) Im[G_EG_M^*]/\sqrt{\tau} \over
       |G_E|^2\sin^2(\theta)/\tau +|G_M|^2(1+\cos^2\theta)}$
 \end{enumerate}
\item One Longitudinally polarized beam particle
 \begin{enumerate}
 \item Recoil polarization $\parallel$ to Baryon:\\
$P_z =  -P_e{2\cos(\theta) |G_M|^2 \over
       |G_E|^2\sin^2(\theta)/\tau +|G_M|^2(1+\cos^2\theta)}$
 \item Recoil polarization in scattering plane  $\perp$ to Baryon:\\
 $P_x =  -P_e{2\sin(\theta) Re[G_EG_M^*]/\sqrt{\tau} \over
       |G_E|^2\sin^2(\theta)/\tau +|G_M|^2(1+\cos^2\theta)}$
 \end{enumerate}
\end{enumerate}

Method 1b allows us to measure the phase difference once the absolute values have
been determined using the Rosenbluth separation (1a). The measurement can be carried out
at all scattering angles simultaneously.

\section{Measuring Polarization}
The polarization of the recoil nucleon can be measured in a polarimeter similar to the one
used recently at JLAB \cite{polarimeter}.  The method is to precisely measure the incoming
trajectory, re-scatter the recoil nucleon on a carbon target, and then measure the
outgoing angle to get the 
angular dependence of the re-scatter. The number of counts $N(\Phi^\prime)$ is given by:
\begin{equation}
N(\theta^\prime,\phi^\prime) = N_o(1 + P\times A_{eff} \sin\Phi^\prime)
\end{equation}
where $P$ is the polarization to be measured, $A_{eff}$ is the effective analyzing power,
and $\Phi^\prime$ is the second scattering angle.  This may require a separate
detector unless we can embed a precise position measuring system within the 
multi-layered shower counter. 
The $\Lambda$ Form Factors could be measured by using the self analyzing power of the
decay. Corrections must be made for the
precession of the Baryon spin  by the field of the vertex magnet.
\subsection{Rates and Errors}
We estimate that  there will be about 200  $N \bar{N}$/day. I assume that we have an
analyzing power similar to that at JLAB which is 0.5 at 230 MeV and 0.1 at a few
GeV.  The probability of scattering $P_s$ in the polarimeter is 0.01 to 0.1 giving
the figure of merit $A_{eff}\sqrt{P_s} \sim 0.07$. The error on the nucleon polarization
is given by $\delta P_s \sim 1./(A\sqrt{P_sN}$ where $N$ is the number of nucleons
going into the polarimeter.
With these parameters, a 100 day run yields $\delta P \sim 0.1$.

\section{Conclusion}
$G_E(s)$ and $G_M(s)$ and their relative phase can be measured at PEP-N using unpolarized
beams and measuring the recoil polarization of the Nucleon.

\end{document}